\documentclass[12pt,british]{article}
\usepackage[T1]{fontenc}
\usepackage[latin9]{inputenc}
\usepackage{geometry}
\usepackage{color}
\usepackage{float}
\usepackage{amstext}
\usepackage{amssymb}
\usepackage{setspace}
\usepackage[authoryear]{natbib}
\usepackage{amssymb}
\usepackage{amsmath}
\usepackage{color}
\usepackage{multirow}
\usepackage{graphicx}
\usepackage{amsfonts}
\usepackage{graphicx}
\usepackage{caption}
\usepackage[hidelinks]{hyperref}

\usepackage[dot]{bibtopic}

\def\bSig\mathbf{\Sigma}

\newcommand{\indep}{\perp\!\!\!\perp}

\makeatletter

\date{}

\newtheorem{definition}{Definition}

\newtheorem{lemma}{Lemma}
\newtheorem{assumption}{Assumption}

\usepackage{babel}

\begin{document}

\title{Propensity Score Matching and Subclassification in \\
\vspace{-0.2cm}
Observational Studies  with Multi-level Treatments}
\vspace{-0.2cm}
\author{
\vspace{-0.25cm}
Shu Yang\footnote{shuyang@hsph.harvard.edu}\\
\vspace{-0.25cm}
Department of Biostatistics, Harvard T. H. Chan School of Public Health\\
\vspace{-0.25cm}
Guido W. Imbens\\
\vspace{-0.25cm}
Graduate School of Business, Stanford University\\
\vspace{-0.25cm}
Zhanglin Cui and Douglas E. Faries\\
\vspace{-0.25cm}
Real World Analytics, Eli Lilly and Company\\
\vspace{-0.25cm}
and Zbigniew Kadziola\\
\vspace{-0.25cm}
Real World Analytics, Eli Lilly and Company
}
\maketitle

\begin{abstract}
In this paper, we develop new methods for estimating average treatment effects in observational studies, in settings with more than two treatment levels, assuming unconfoundedness given pre-treatment variables. We emphasize propensity score subclassification and matching methods which have been among the most popular methods in the binary treatment literature. Whereas the literature has suggested that these particular propensity-based methods   do not naturally extend to the multi-level treatment case, we show, using the concept of weak unconfoundedness and the notion of the generalized propensity score,
 that adjusting for a scalar function of the pre-treatment variables
removes all biases associated with  observed pre-treatment variables. We apply the proposed methods to an analysis of the effect of treatments for fibromyalgia. We also carry out a simulation study to assess the finite sample performance of the methods relative to previously proposed methods.
\vspace{0.1in}
\end{abstract}

%



\maketitle


%

\section{Introduction}
\label{sec:intro}

There is an extensive theoretical and empirical literature on estimating average causal effects of binary treatments in observational studies based on the  assumption of unconfoundedness or ignorable treatment assignment. Under this assumption
differences in outcomes for units with different treatment levels, but the same values for pre-treatment variables, can be interpreted as estimates of causal effects. Much of the literature builds on the seminal paper by \citet{RR:1983} (RR83 from here on) which clarified the central role of the propensity score (the conditional probability of receiving the treatment given  the pre-treatment variables or covariates) in analyses of causal effects in such settings, and which    proposed a number of  widely used estimators. 
See \citet{IR:2015} for a textbook treatment.

Although important for empirical practice, much less theoretical work has been done on the setting with more than two treatment levels (exceptions include \citealp{I:2000}; {\citealp{R:2000}}; \citealp{L:2001, F:2003, HI:2004, IV:2004, CF:2009, CA:2010, C:2010, M:2013, R:2013}).  
Because in  settings with multi-level treatments there is no scalar function of the covariates that has all the  properties that RR83 presents for the propensity score in the binary treatment case,  it has been claimed  that there is no natural analogue to matching and subclassification on the  propensity score
(\citealp{I:2000, L:2001, R:2013}).

In the main contribution of the current paper we show that, contrary to these claims, the essence of the results in RR83 generalizes to the setting with multi-level treatments. In particular, we develop methods for matching and subclassification on scalar functions of the covariates that are valid irrespective of the number of distinct levels for the treatment. The key insight is that we do {not} construct sets  of units with the balancing property that within these sets the treatment level is independent of the covariates. Doing so would
require adjusting for the vector of propensity scores with length equal to the number of treatment levels minus one.
Instead we focus on estimating the average of the potential outcomes separately for each treatment level, which requires adjusting only for the probability of receiving that particular level of the treatment.  
This insight allows us to extend some
of the most widely used methods for the binary treatment case to the multi-level treatment case without giving up the dimension reducing property of the propensity score. We provide some simulation evidence that demonstrates the relevance of  concerns with the previously proposed estimators and the promise  of the new methods.


\section{Set Up}
\label{sec:setup}

Following \citet{R:1974} and RR83
we use the 
 potential outcome set up, 
 generalized to the case with more than two, unordered, levels for the treatment in \citet{I:2000}, \citet{L:2001},  \citet{IV:2004}, and \citet{C:2010}. The treatment is denoted by $W_i\in \mathbb{W}=\{1,\ldots,T\}$. In the standard binary treatment case $T=2$, the two treatments are often labeled treatment and control.  For each unit $i$ there are $T$ potential outcomes, one for each treatment level, denoted by $Y_i (w)$, for $w\in \mathbb{W}$. Implicitly in this notation is the assumption that there is no interference between units and no versions of each treatment level
(the stable-unit-treatment-value assumption, or sutva, \citealp{R:1978}). The  observed outcome for unit $i$ is the potential outcome corresponding to the treatment received:
\begin{equation}
\nonumber 
Y_i^{\rm obs} = Y_i (W_i).
\end{equation}
We also observe a vector-valued covariate or pre-treatment variable, denoted by $X_i$. These pre-treatment variables are known a priori not to be affected by the treatment, typically measured prior to the determination of the treatment level,  and so there are no multiple versions of these covariates corresponding to the different levels of the treatment. These pretreatment variables  may include fixed attributes of the units, or measurements prior to the treatment assignment that predict the outcome, for example prior health status. 
Although we do not stress this in the notation, there is implicitly a temporal aspect to the study with three stages: first, the pre-treatment variables are measured, or at least they assume their values prior to the assignment of the treatment; second, the treatment is assigned or selected; third, the outcome assumes its value and is measured.

We assume the  sequence $(W_{1},X_{1},Y_{1}(1),\ldots,Y_{1}(T)),\ldots,(W_N,X_N,Y_{N}(1),$ $\ldots,Y_{N}(T))$ with the potential outcomes is i.i.d.,  so that the sequence of realized values $(W_1,X_1,Y^{\rm obs}_1),\ldots,(W_N,$ $X_N,Y_N^{\rm obs})$ is also i.i.d. 

Following the literature (e.g., RR83), we focus on average treatment effects as the causal estimands.
This is less restrictive than it may appear at first because we can first take transformations of the outcomes and pre-treatment variables.  
For the comparison between treatment levels $w$ and $w'$, the average effect is
\begin{equation}
\label{2.1}
\tau (w,w')
=  \mathbb{E} [Y_i (w') - Y_i (w)]. 
\end{equation}
The expectation is taken with respect to the same population (called the target population, \citealp{F:2004}) for different treatment-level pairs $(w,w')$. 
Some researchers, when analyzing data with multi-level treatments, have used conventional methods for comparing two treatment levels at a time. Often such analyses use only information on units exposed to one of those two treatment levels, which would lead to estimates of $\mathbb{E}[Y_i (w')-Y_i (w)\mid W_i\in \{w,w'\}]$. If the subpopulation of units with treatment levels $w$ or $w'$ is different in terms of potential outcome distributions, these estimands  are generally different from the $\tau(w,w')$ defined in (\ref{2.1}), because the latter do not condition on $W_i\in \{w,w'\}$. As a result such binary-comparison analyses make it difficult to compare 
$\mathbb{E}[Y_i (w')-Y_i (w)\mid W_i\in \{w,w'\}]$ and
$\mathbb{E}[Y_i (w'')-Y_i (w)\mid W_i\in \{w,w''\}]$ because they refer to different populations.

In this paper we mainly focus on the case where the different treatment levels are qualitatively distinct. In that cases the interest is typically in average effects of the form $\tau(w,w')$. In other cases, however, the treatment levels may measure the quantity of a dose. In such cases the researcher may be interested in weighted combinations of average effects. For example, one might be interested in $\sum_{w=1}^{T-1} \lambda_w \tau(w,w+1)$, with the weights adding up to one, which would correspond to a weighted average of unit increases in the dose. One advantage of such estimands is that their variance may be lower than that for particular $\tau(w,w')$. In this case there may also be particular interest in $\tau(1,T)$, the effect of the maximum dose. Our results also apply to all such estimands.


\section{Weak and Strong Unconfoundedness and the Generalized Propensity Score}
\label{sec:ws}

Our focus is on observational studies where assignment to  treatment is not completely random. Instead, following a large strand of the observational studies literature, we assume that assignment to treatment is unconfounded so that,  within subpopulations that are homogenous in observed pre-treatment variables, assignment to treatment is as good as random.
This is strictly weaker than complete randomization by allowing for general associations between the treatment level and the pre-treatment variables. 

\subsection{The generalized propensity score}

In this section we discuss the generalization of the notion of the propensity score, introduced in the causality literature by RR83 for the binary treatment case, to our setting with multi-level treatments.
In the binary treatment case RR83 defines the propensity score as the conditional probability of receiving the active treatment rather than the control treatment, $p(x)=pr(W_i=1\mid X_i=x)$. Here we generalize that to the multi-level treatment case, following \citet{I:2000}:
\begin{definition}
(Generalized Propensity Score) The generalized propensity score is the conditional probability of receiving each treatment level:
\begin{equation}
\nonumber p(w\mid x) = pr(W_i=w\mid X_i=x). 
\end{equation}
\end{definition}

\subsection{Overlap}
Before formally discussing unconfoundedness assumptions, let us assume that there is overlap in the covariate distributions:
\begin{assumption}
(Overlap) For all values of $x$ the probability of receiving any level of the treatment is positive:
\begin{equation}
\nonumber p(w\mid x) >0 \quad \text{ for all } w, x.
\end{equation}
\end{assumption}
Without this assumption there will be values of $x$ for which we cannot estimate the average effect of some treatments relative to others without relying on extrapolation.  In  Section 6 we discuss  methods 
for constructing a subsample with better overlap for cases where this assumption is (close to) violated.

\subsection{Strong unconfoundedness}
We start by generalizing the conventional RR83 version of the unconfoundedness assumption to the case with multi-level treatments. We refer to this as strong unconfoundedness to distinguish it from the weaker condition of weak unconfoundedness.
\begin{definition}
(Strong Unconfoundedness) The assignment mechanism is strongly unconfounded if
\begin{equation}
\nonumber W_i\ \indep\ \Bigl(Y_i(1),\ldots, Y_i(T)\Bigr)\ \mid\  X_i.
\end{equation}
\label{strong}
\end{definition}
\vspace{-0.8cm}
Here we use the  $\indep$ notation introduced by \citet{D:1979}  to denote (conditional) independence. 

The assumption of strong unconfoundedness has no testable implications. 
In a particular application the assumption 
 is a substantive one, and often a  controversial one. Often it can be made more plausible by collecting detailed information at baseline on characteristics of the units that are related to treatment and outcome. As a result the dimension of $X_i$ may be high.

One implication of strong unconfoundedness is the following extension of the propensity score result in RR83 to the multi-level treatment case:
\begin{lemma}
\label{lemma1}
(RR83) Suppose the assignment mechanism is strongly unconfounded. Then
\begin{equation}\nonumber
 W_i\ \amalg\ \Bigl(Y_i(1),\ldots, Y_i (T)\Bigr)  \left|  \Bigl(p(1\mid X_i) ,\ldots, p(T-1 \mid X_i )\Bigr)\right..
\end{equation}
\end{lemma}
Because $\sum_{w=1}^T p(w|x)=1$, it follows that $p(T|x)$ is a linear combination of $p(1| x),\ldots, p(T-1| x)$, and so we do not need to include $p(T| x)$ in the conditioning set. If there are two levels of the treatment, the result in the lemma reduces to
the result in RR83. As pointed out in \citet{I:2000} and \citet{R:2013},  the dimension reduction property of the lemma depends on the number of distinct levels for the treatment, and therefore the result is less useful in settings with a substantial number of treatment levels. The problem is that without additional assumptions there is in general no scalar function $b(x)$ of the covariates such that $W_i \indep (Y_i(1),\ldots, Y_i (T))  \mid  b(X_i)$, suggesting that the advantages of the propensity score approach do not carry over to the multi-level treatment case. \cite{Jo:1999,Lu:2001,IV:2004,Z:2005}
 discuss additional assumptions under which functions $b(\cdot)$ exist with this property and whose dimension is lower than $T-1$. In particular, \cite{Lu:2001} assume that a scalar balancing function $b(\cdot)$ exists and propose a matching estimator based on $b(\cdot)$, and \cite{Z:2005} propose a subclassification estimator under this assumption. Nevertheless,   
in general such functional form assumptions may be controversial. 



\subsection{Weak unconfoundedness}

We improve the dimension reduction property of the generalized propensity score by weakening the requirement of strong unconfoundedness condition to weak unconfoundedness. Define the $T$ indicator variables $D_i (w)\in\{0,1\}$:
\begin{equation}
\nonumber
D_i (w)=\begin{cases}
1  ~ \text{ if  } W_i = w, \\
0  ~ \text{ otherwise. } 
\end{cases}
\end{equation}
In terms of these indicator variables strong unconfoundedness is equivalent to
\begin{equation}
\nonumber
\Bigl(D_i(1),\ldots,D_i(T-1)\Bigr)\  \indep\  \Bigl(Y_i(1),\ldots,Y_i(T)\Bigr)\ \mid\  X_i.   
\end{equation}
Now we can formulate the weak unconfoundedness notion, introduced in \citet{I:2000}.
\begin{definition}
(Weak unconfoundedness) The assignment mechanism is weakly unconfounded if
for all   $w\in\mathbb{W}$, 
\begin{equation}
\nonumber
D_i(w)\  \indep\  Y_i(w)\ \mid\  X_i.   
\end{equation}
\end{definition}
Although formally it is obviously weaker, we do not wish to argue that  weak unconfoundedness is substantively weaker than strong unconfoundedness. In fact neither have testable implications, and there appear to be no interesting estimands that are identified under the stronger assumption but not under the weaker assumption. Rather, the two key insights, and the motivation for distinguishing between the two unconfoundedness assumptions, are, one, that, as shown in Lemma \ref{lemma2} below, weak unconfoundedness is preserved if we condition on a scalar function of the pretreatment variables,  whereas preserving strong unconfoundedness requires conditioning on a set of $T-1$ functions of the  pre-treatment variables, as shown in Lemma \ref{lemma1}, and two, that weak unconfoundedness is sufficient for identifying average causal effects, as formalized in Lemma \ref{lemma3} below.
\begin{lemma}\label{lemma2}
(Weak Unconfoundedness) Suppose the assignment mechanism is weakly unconfounded. Then for all $w\in\mathbb{W}$, 
\begin{equation}
\nonumber
D_i(w)  \ \indep\  Y_i(w)\ \mid\  p(w|X_i).
\end{equation}
\end{lemma}
\begin{lemma}\label{lemma3}
(Average Causal Effects Under Weak Unconfoundedness) Suppose the assignment mechanism is weakly unconfounded. Then
\begin{equation}
\nonumber
\mathbb  E[Y_i(w') - Y_i(w)]
=  \mathbb{E}\Bigl[  \mathbb{E}[Y_i^{\rm obs}\mid W_i=w',p(w'\mid X_i)]\Bigr]-  \mathbb{E}\Bigl[  \mathbb{E}[Y_i^{\rm obs}\mid W_i=w, p(w\mid X_i)]\Bigr].
\end{equation}
\end{lemma}
Lemma \ref{lemma3} is the key result. 
For its interpretation it is useful to compare it to the standard result under strong unconfoundedness.
Under the strong unconfoundedness assumption we create subpopulations where we can simultaneously compare units with all different levels of the treatment, leading to
\begin{equation*}
\mathbb  E[Y_i(w') - Y_i(w)]\end{equation*}
\begin{equation*}
=  \mathbb{E}\Bigl[  \mathbb{E}[Y_i^{\rm obs}\mid W_i=w',p(1\mid X_i),\ldots,p(T-1|X_i)]\Bigr]-  \mathbb{E}\Bigl[  \mathbb{E}[Y_i^{\rm obs}\mid W_i=w,p(1\mid X_i), \ldots,p(T-1\mid X_i)]\Bigr]
\end{equation*}
\begin{equation*}
=\mathbb{E}\Bigl[  \mathbb{E}[Y_i(w')-Y_i(w)\mid p(1\mid X_i),\ldots,p(T-1|X_i)]\Bigr]
\end{equation*}
To allow for comparisons of all treatments these subpopulations were defined by common values for the full set of $T-1$ propensity scores $(p(1\mid X_i),\ldots,p(T-1\mid X_i))$. Under weak unconfoundedness we do not, and in fact cannot,  construct such subpopulations. However, in order to estimate the average effect $E[Y_i(w') - Y_i(w)]$ it is not necessary to do so. Instead, we construct, for each treatment level $w$ separately, subpopulations where we can estimate the average value of the potential outcomes, but only for that single treatment level. For treatment level $w$ these subpopulations are defined by the value of a single score, $p(w|X_i)$, leading to the equality
\begin{equation}
\nonumber
\mathbb  E[Y_i(w)]
=    \mathbb{E}\Bigl[  \mathbb{E}[Y_i^{\rm obs}\mid W_i=w, p(w\mid X_i)]\Bigr].
\end{equation}
 That difference in focus allows us to reduce the dimension of the conditioning variable to a scalar, irrespective of the number of treatment levels.


\section{Matching}
\label{sec:CM}

In this section we discuss matching methods. First we discuss conventional matching on the full set of pre-treatment variables. This is not a new method, but it will be useful to contrast with the proposed methods. Then we discuss how the generalized propensity score can be used to develop a new matching estimator that matches only on a scalar function of the pre-treatment variables.

\subsection{Matching}
\label{sec:CM}

\cite{F:2004b} demonstrates covariate matching in multi-level treatments. Here we focus on nearest neighbor matching. Other modifications include multiple nearest neighbors matching, kernel matching and so forth. Reviews of  general matching methods can be found in \cite{I:2004, IR:2015, H:2013}.
Define the covariate matching function $m_{\rm cov} :   \mathbb{W} \times  \mathbb{X} \mapsto \{1,\ldots,N\}$ as the index for the unit with treatment level $w$ that is closest to $x$ in terms of covariates (ignoring ties):
\begin{equation}
\nonumber
m_{\rm cov}(w,x) = \arg \min\limits_{j: W_j = w} ||X_j - x||.
\end{equation}
Here we use $||\cdot||$ to denote a generic metric. In practice one would typically use the Mahalabonis metric, where $||x-x'||=\{(x-x' )^T V^{-1} (x-x')\}^{1/2}$, with $V=\sum_i(X_i-\overline{X})(X_i-\overline{X})^T/N$, and $\overline{X}=\sum_i X_i /N$. Note that the set of indices we search over includes all units, including unit $i$ itself, so that for all $i$, $m_{\rm cov}(W_i,X_i) = i$. Given the covariate matching function $m_{\rm cov}(w,x)$ the potential outcomes for unit $i$ are imputed as
\begin{equation}
\nonumber
\hat{Y}_i(w) = Y_{m_{\rm cov}(w,X_i)}^{\rm obs},
\end{equation}
for $w=1, \ldots, T$.
Now we estimate 
 $\tau(w,w')$ as
\begin{equation}\label{twee}
\hat{\tau}_{\rm cov}(w,w') 
                       \ = N^{-1} \sum\limits_{i=1}^N \left(Y_{m_{\rm cov}(w',X_i)}^{\rm obs}
-Y_{m_{\rm cov}(w,X_i)}^{\rm obs}\right).\end{equation}
Note that to estimate $\tau(w,w')$ we  impute potential outcomes $Y_i(w)$ and $Y_i(w')$ even for units who did not receive either treatment level $w$ or $w'$. This ensures comparability of average treatment effects for different pairs of treatments.


\subsection{Matching on the Generalized Propensity Score}
\label{sec:mg}

Just as in the binary treatment setting, matching on all covariates is not an attractive procedure in the multi-level treatment setting if the number of covariates is substantial ({\it e.g.},  \citealp{AI:2006, IR:2015, IV:2004}). In the binary treatment case RR83 proposed matching on the propensity score to reduce the dimensionality of the matching problem. If $p(1|x)$ is the Rosenbaum-Rubin propensity score, the matching function for the binary treatment case would  be
\begin{equation}
m_{\rm ps}^{\rm binary}(w,p) = \arg \min\limits_{j: W_j = w} ||p(1|X_j) - p||.
\end{equation}
One could generalize that to the multi-level treatment case by matching on the full set of scores, leading to
\begin{equation}
m_{\rm gps}^{\rm multilvl}(w,p_1,\ldots,p_{T-1}) = \arg \min\limits_{j: W_j = w} \left\|\left(\begin{array}{c}
p(1|X_j) - p_1 \\
\vdots\\
p(T-1|X_j) - p_{T-1} \end{array}\right)
\right\|.
\end{equation}

Here we generalize this to the case with multi-level treatments in a way that allows for a scalar matching variable. In this case matching is conceptually quite different from matching on covariates.
We separate the estimation of $\tau(w,w')=  \mathbb E[Y_i (w')]-\mathbb E[Y_i(w)]$ into the two terms. First we focus on the problem of estimating $  \mathbb E[Y_i(w)]$.
Define the generalized propensity score matching function as
\begin{equation}
m_{\rm gps}(w,p) = \arg \min\limits_{j: W_j = w} ||p(w| X_j) - p||.
\end{equation}
Here the treatment level $w$ enters into the matching function not only by limiting the set of potential matches to the set of units with $W_j=w$, but also in the function of the covariates that is being matched, $p(w|X_j)$. In  covariate and conventional propensity score  matching  the treatment level only affects the set of potential matches.

Given the generalized propensity score matching function we  impute $Y_i(w)$ as
\begin{equation}
\nonumber
\hat{Y}_i(w) = Y_{m_{\rm gps}(w,p(w\mid X_i))}^{\rm obs}.
\end{equation}
The average effect is estimated as
\begin{equation}\label{drie}
\hat{\tau}_{\rm gps}(w,w')  =  N^{-1} \sum\limits_{i=1}^N \left(Y_{m_{\rm gps}(w',p(w'\mid X_i))}^{\rm obs} -  Y_{m_{\rm gps}(w,p(w\mid X_i))}^{\rm obs}\right).
\end{equation}
Note that the difference $Y_{m_{\rm gps}(w',p(w'\mid X_i))}^{\rm obs} -  Y_{m_{\rm gps}(w,p(w\mid X_i))}^{\rm obs} $ in (\ref{drie}) is {\it not} generally an estimate of an average causal effect, whereas in (\ref{twee}) the difference $Y_{m_{\rm cov}(w',X_i)}^{\rm obs} - Y_{m_{\rm cov}(w,X_i)}^{\rm obs}$ {\it is} an estimate of the average causal effect $\mathbb  E[Y_i (w')-Y_i (w)\mid X_i]$.  In the binary treatment case this distinction between covariate and propensity score matching does not matter: in that case there are only two values for $w$, $w = 1,2$, so that $p(1\mid x) = 1 - p(2\mid x)$, and therefore matching on $p(1\mid x)$ is the same as matching on both $p(1\mid x)$ and $p(2\mid x)$, and the same as matching on $p(2\mid x)$. 

{{In Web Appendix, we provide mathematical details for inference and show that under mild regularity conditions, the matching estimator based on the generalized propensity score or the estimated generalized propensity score is asymptotically normally distributed.}}

\section{Subclassification on the Generalized Propensity Score}
\label{sec:subclassification}

In the binary treatment literature, a common alternative to matching is subclassification or stratification on the propensity score, originally proposed by RR83. To put our proposed methods for the multi-level treatment case in perspective, let us briefly summarize their approach for the binary treatment case in our current notation to show why it does not directly extend to the multivalued treatment case. Divide the sample into a number of subclasses by the value of the propensity score $p(1|x)$. Based on Cochran (1968) who shows that this removes much of the bias, researchers often use five subclasses. To be specific, let $q_j^{p(1|x)}$ be the $j$th quintile of the empirical distribution of $p(1|X_i)$, for $j = 1,
\ldots, 4$, and define $q_0^{p(1|x)}= 0$ and $q_5^{p(1|x)}=1$. Then we construct the five subclasses, based on the propensity score being between $q_{j-1}^{p(1|x)}$ and $q_j^{p(1|x)}$. For subclass $j$ one can estimate the average causal effect of treatment $1$ versus treatment $2$ as 
\begin{equation}
\nonumber
\tau_j(1,2) = \frac{1}{N_{j2}}\sum\limits_{i:q_{j-1}^{p(1|x)}<p(1|X_i)\leq q_j^{p(1|x)}, W_i=2} Y_i^{\rm obs} - \frac{1}{N_{j1}}\sum\limits_{i:q_{j-1}^{p(1|x)}<p(1|X_i)\leq q_j^{p(1|x)}, W_i=1} Y_i^{\rm obs},
\end{equation}
where $N_{jw}$ is the number of units in subclass $j$ with treatment level $w$. The overall average treatment effect is then estimated by averaging over the subclasses:
\begin{equation}
\nonumber
\hat{\tau}(1,2) = \sum\limits_{j=1}^5 \frac{N_{j1}+N_{j2}}{N}\cdot \hat{\tau}_j(1,2).
\end{equation}
Because all $N_{j1}+N_{j2}$ are close to equal, at most differ by $1$, (assuming there are no ties) this is essentially a simple arithmetic mean of the $J$ estimates $\hat\tau_j(1,2)$. 

Now  consider the multi-level treatment case. We are interested in $\tau(w,w')$ for some pair of treatment levels $w$ and $w'$. Again, and this is a cornerstone of our approach, we write this as a difference of two expectations, $\tau(w,w')=  \mathbb E[Y_i(w')]- \mathbb E[Y_i(w)]$ and separately estimate the two terms $\mathbb  E[Y_i(w')]$ and $\mathbb E[Y_i(w)]$.
To estimate the second term, $ \mathbb E[Y_i(w)]$ we construct subclasses or strata based on $p(w|x)$. Let $q_j^{p(w\mid x)}$ be the quintiles of $p(w\mid X_i)$ in the sample. Then the average value of $Y_i(w)$ in subclass $j$ is estimated as 
\begin{equation}
\nonumber
\hat{\mu}_{jw}= \frac{1}{N_{jw}}\sum\limits_{i:q_{j-1}^{p(w\mid x)}<p(w\mid X_i)\leq q_j^{p(w\mid x)}, W_i=w} Y_i^{\rm obs},
\end{equation}
where $N_{jw}$ is the number of units with $q_{j-1}^{p(w\mid x)}<p(w\mid X_i)\leq q_j^{p(w\mid x)}$ and $W_i=w$.
The overall average of $Y_i(w)$ is then estimated as
\begin{equation}
\nonumber
\hat{\mathbb  E}[Y_i(w)] = \sum\limits_{j=1}^5 \frac{N_{w}}{N}\cdot \hat{\mu}_{jw}.
\end{equation}

The key is that, in contrast to what is done in the binary treatment case, we do not construct subclasses 
defined by similar values for the $T-1$ propensity scores 
such that we can estimate causal effects within the subclasses. Instead we construct subclasses     {defined by similar values for a single propensity score at a particular treatment level} so that we can estimate the average potential outcome for that treatment level within the subclasses, 
and we do so separately for each treatment level, with different subclasses for each treatment level. 
In the binary treatment case this amounts to the same thing because the two propensity scores $p(1|x)$ and $p(2|x)$ are linearly related, but in the multi-level case these two approaches are different.

\section{Assessing And Ensuring Overlap}

\subsection{Assessing balance}

We focus on assessing balance in the covariate distributions in terms of the propensity score as well as directly in terms of the covariates, following the discussion in \citet{IR:2015} for the binary case.
For each treatment level $w$, we calculate the average values for each component of the covariate vectors and their corresponding sample variances:
\begin{equation}
\nonumber
\overline{X}_{w} = N_w^{-1} \sum\limits_{i:W_i=w} X_i, \quad \text{ and } {S}_{X,w}^2 = (N_w-1)^{-1} \sum\limits_{i:W_i=w} (X_i-\overline{X}_w)^2.
\end{equation}
Define also for each treatment level the average value of the covariates for units with a treatment level different from $w$ and the average variance:
\begin{equation}
\nonumber
\overline{X}_{\overline{w}} = (N-N_w)^{-1} \sum\limits_{i:W_i\neq w} X_i, \quad \text{ and } {S}_{X\mid W}^2 = T^{-1} \sum\limits_{i=1}^T S_{X,w}^2.
\end{equation}
respectively. The first approach to assessing the covariate balance is to inspect the normalized differences for each covariate and each treatment level:
\begin{equation}
nd_w^{COV} = (\overline{X}_w - \overline{X}_{\overline{w}})/S_{X\mid W}
\end{equation}
We can also assess balance by looking at the generalized propensity score. For each treatment level we can calculate the normalized difference for the generalized propensity score for that treatment level:
\begin{equation}
nd_w^{GPS} = \left(\overline{p(w\mid X)}_w - \overline{p(w\mid X)}_{\overline{w}}\right)/S_{p(w\mid X)\mid W}
\end{equation}
where $\overline{p(w\mid X)}_w = N_w^{-1} \sum\limits_{i:W_i=w}p(w\mid X_i)$  and $\overline{p(w\mid X)}_{\overline{w}}=(N-N_w)^{-1}\sum\limits_{i:W_i\neq w} p(w\mid X_i)$.
Finally, one may wish to plot a histogram of $p(w\mid X_i)$ for the $N_w$ units for $W_i=w$ and a histogram of $p(w\mid X_i)$ for the $N-N_w$ units with $W_i\neq w$ in the same figure.

\subsection{Improving Overlap}

In many applications there are regions of the covariate space with low values for the probability of receiving one of the treatments. This is likely in the setting with a binary treatment, but even more likely
to be an issue in settings with many treatment levels. Of note, lack of overlap affects the credibility of all methods attempting to estimate all pairwise average causal effects from the common population.
In that case we may wish to modify the estimands to average only over the part of the covariate space with all treatment probabilities away from zero. The question is how to choose the set of covariates with overlap. 
For the binary treatment case, \citet{C:2009} proposed a method for improving overlap by trimming the sample. Specifically they suggest dropping units from the analysis with low and high values of the propensity score. The threshold for dropping units is based on minimizing the variance of the estimated average treatment effect on the trimmed sample. 
    By trimming the sample, this method generally alters the estimand to the so-called feasible estimand, by changing the reference population.
Using the feasible estimand is widely recommended in the literature, as long as we are careful to characterize the resulting quantity of interest.
Here we generalize the \citet{C:2009} approach to the multi-level treatment case.
We focus on average treatment effects for subsets of the covariate space. Formally, define the conditional average treatment effect:
\begin{equation}
\nonumber
\tau(w,w'| \mathbb{C}) = \mathbb  E[Y_i(w') - Y_i(w)\mid X_i \in  \mathbb{C}].
\end{equation}
The semiparametric efficiency bound for $\tau(w,w'\mid \mathbb{C})$ is, building on the work by Hahn (1998) and Hirano, Imbens, and Ridder (2003), under homoskedasticity and constant treatment effects,
\begin{equation}
\nonumber
\mathbb V(w,w'|\mathbb{C}) =  \frac{\sigma^2}{pr(X_i\in  \mathbb{C})}\mathbb E \left[
\frac{1}{p(w | X_i)}+
\frac{1}{p(w'| X_i)}\mid X_i\in  \mathbb{C} \right].
\end{equation}
In the binary case \citet{C:2009} proposed choosing $  \mathbb{C}$ to minimize $\mathbb V(w,w'\mid   \mathbb{C})$, which leads  to dropping units with $p(1\mid X_i)\leq \alpha$ and units with $p(1\mid X_i)\geq 1-\alpha$, with $\alpha$ an estimable function of the marginal distribution of the propensity score. 

For the multi-level treatment case we suggest focusing on the subset of the covariate space $  \mathbb{C}$ that minimizes
\begin{equation}
\nonumber
\overline{  \mathbb V}(  \mathbb{C}) = \sum\limits_{w,w'}   \mathbb V(w,w'\mid   \mathbb{C}) =\frac{2\sigma^2}{pr(X_i\in  \mathbb{C})}
\mathbb E \left[\sum\limits_{w=1}^T\frac{1}{p(w\mid X_i)}\mid X_i\in   \mathbb{C} \right].
\end{equation}
Under homoskedasticity and a constant treatment effect this will lead to a set $\mathbb{C}$ of the form
\begin{equation}
\nonumber
  \mathbb{C} = \left\{X_i\in\mathbb X \left| \sum\limits_{w=1}^T\frac{1}{p(w\mid X_i)}\leq \lambda \right.\right\},
\end{equation}
where the threshold 
$\lambda$ satisfies \begin{equation} \nonumber \lambda \leq \frac{2}{pr\Big(\sum\limits_{w=1}^T(p(w\mid X_i))^{-1}\leq \lambda\Big)}\mathbb  E \left[\sum\limits_{w=1}^T\frac{1}{p(w\mid X_i)}\left| \sum\limits_{w=1}^T\frac{1}{p(w\mid X_i)}\leq \lambda\right. \right].\end{equation}
To implement the trimming method in practice in the multi-level treatment case we replace the expectation by an average and then find the largest $\lambda$ that satisfies the inequality.

\section{A Simulation Study}

In this section we assess the performance of the two new estimators
in cases of multi-level treatments (matching on the generalized propensity
score, GPSM, and subclassification on the generalized propensity score,
GPSS) in a Monte Carlo study relative to five previously proposed
estimators, first the simple difference in average outcomes (DIF)
by treatment status, second pairwise propensity score matching (PPSM)
that compares two treatment levels at a time using the binary propensity
score matching on the units exposed to one of those two treatment
levels, third the estimator based on matching on the set of $T-1$
propensity score set (PSSM), fourth the estimator based on weighting,
and fifth, matching on all covariates (COV). In the binary treatment
and missing data case previous simulations have found that weighting
estimators can have high variability, \textit{e.g.,} \citet{KS:2007}
and \citet{GF:2010}, especially if the probabilities are close to
zero. \citet{F:2004} found that the weighting estimator was inferior
to pairwise matching estimators in terms of root mean squared error.
This is even more likely to be a concern in settings with multiple
treatment levels than in the binary treatment case because, with the
probabilities for the $T$ treatment levels adding up to one, with
$T$ large some probabilities are likely to be close to zero. Because
in the binary treatment case it has been found that matching on high-dimensional
covariates is not practical for commonly found sample sizes (\textit{e.g.,}
the theoretical results in \citealp{AI:2006}), it is likely that
in settings with many treatment levels matching on all scores is not
effective either. These results motivate us to compare the seven estimators
in settings with a large number of treatment levels, and where some
of the treatment levels have low probability for some covariate values.
In the simulations we focus on two designs, one with three treatment
levels and one with six treatment levels, and both with six covariates.

In the first design with three treatment levels the covariates $X_{1i},X_{2i}$,
and $X_{3i}$ are multivariate normal with means zero, variances of
$(2,1,1)$ and covariances of $(1,-1,-0.5)$; $X_{4i}\sim U[-3,3];X_{5i}\sim\chi_{1}^{2}$;
and $X_{6i}\sim{\rm {Bernoulli}(0.5)}$, with $X_{i}^{T}=(1,X_{1i},X_{2i},X_{3i},X_{4i},X_{5i},X_{6i})$.
The three treatment groups are formed using multinomial regression
model 
\[
(D_{i}(1),D_{i}(2),D_{i}(3))\sim{\rm {Multinom}}(p(1|X_{i}),p(2|X_{i}),p(3|X_{i})),
\]
where $D_{i}(w)$ is the treatment indicator, i.e. $D_{i}(w)=1$,
if the unit $i$ belongs to treatment $w$, and $p(w|X_{i})=\exp(X_{i}^{T}\beta_{w})/\sum_{w'=1}^{3}\exp(X_{i}^{T}\beta_{w'})$,
$\beta_{1}^{T}=(0,0,0,0,0,0,0)$, $\beta_{2}^{T}=0.7\times(0,1,1,1,-1,1,1)$,
and $\beta_{3}^{T}=0.4\times(0,1,1,1,1,1,1)$. The outcome design
is $Y_{i}(w)=X_{i}^{T}\gamma_{w}+\eta_{i}$ with $\eta_{i}\sim N(0,1)$,
$\gamma_{1}^{T}=(-1.5,1,1,1,1,1,1)$, $\gamma_{2}^{T}=(-3,2,3,1,2,2,2)$,
and $\gamma_{3}^{T}=(1.5,3,1,2,$ $-1,-1,-1)$. The sample sizes are
$N_{w}=500$, for $w=1,2,3$.

We compare the seven estimators over $1000$ datasets. The generalized
propensity scores are estimated using multinomial logistic regression
model with all covariates entering the model linearly. $95\%$ confidence
intervals for point estimates were calculated using: (a) $2.5$ and
$97.5$ percentiles from $1000$ bootstrap samples for DIF, GPSS, and weighting;
(b) point estimate $\pm1.96\times(\text{variance})^{1/2}$ for \citet{AI:2006}
variance estimator for COV and PSSM; and for \citet{AI:2012} variance
estimator for PPSM and GPSM, which takes into account the uncertainty
of the matching procedure and the uncertainty of estimating generalized
propensity scores, as in Web Appendix.

\begin{table}[!p]
\centering
\caption{Simulation Results, Design I. Estimators: (1) DIF, simple difference in outcomes for units with different treatment levels; (2)PPSM: pairwise comparison using binary propensity score matching; (3) PSSM: matching on the propensity score set; (4) W: weighting estimator; (5) COV: matching on all covariates; (6) GPSM: matching on the generalized propensity score; (7) GPSS: stratification on the generalized propensity score. Variance estimators: (1) bootstrapping variance estimator for DIF, GPSS, and W; (2) Abadie and Imbens (2006) variance estimator for COV matching and PSSM; (3) Abadie and Imbens (2012) variance estimator for PPSM and GPSM.}
\begin{tabular}{p{1.8cm} p{0.9cm} p{0.9cm} p{0.9cm} p{0.8cm} p{0.8cm} p{0.8cm} p{0.8cm} p{0.8cm} p{0.8cm}}
\hline
\multirow{2}{*}{ } & \multicolumn{3}{c}{Bias} & \multicolumn{3}{c}{RMSE  }  & \multicolumn{3}{c}{ Coverage $95\%$ CI}  \\
 & $\tau(1,2)$  & $\tau(1,3)$  & $\tau(2,3)$  & $\tau(1,2)$  & $\tau(1,3)$  & $\tau(2,3)$  & $\tau(1,2)$  & $\tau(1,3)$  & $\tau(2,3)$ \\
\hline
DIF  & 1.34  & 0.57  & -0.77  & 1.38 & 0.60 & 0.83 & 0.01  & 0.26  & 0.41 \\
PPSM & $-0.6$ & $-1.1$ & $-0.8$ & 0.77 & 1.16 & 0.91 & 0.80 & 0.001 & 0.74\\
PSSM  & 0.21 & 0.19 & -0.02  & 0.27 & 0.33 & 0.29 & 0.90 & 0.92 & 0.98\\
W  & 0.05  & 0.02  & -0.03  & 0.55 & 0.43 & 0.53 & 0.91  & 0.97  & 0.94\\
COV  & 0.29  & 0.19  & -0.11  & 0.33 & 0.22 & 0.20 & 0.75  & 0.88  & 0.99 \\
GPSM  & 0.14  & 0.04  & -0.10  & 0.56 & 0.36 & 0.61 & 0.95  & 0.95  & 0.95 \\
GPSS  & 0.31  & 0.05  & -0.27  & 0.53 & 0.24 & 0.54 & 0.91  & 0.99  & 0.94 \\
\hline
\end{tabular}
\label{table:tab7}
\end{table}

Table 1 presents the bias, root mean squared error (RMSE) and coverage
of $95\%$ confidence intervals. DIF shows that there are substantial
biases associated with the covariates. PPSM compares two treatment
levels at a time using the units exposed to one of those two treatment
levels, which focuses on different populations of inference each time.
This leads to inconsistency for simultaneous comparison of treatment
levels. One implication is that $\hat{\tau}(1,2)+\hat{\tau}(2,3)+\hat{\tau}(3,1)\neq0$.
Even with only three treatment levels, and so only two propensity
scores to match on, PSSM did not control the bias well. The four remaining
procedures, including COV, GPSM and GPSS, and weighting, do a fairly
good job of reducing the bias for all average treatment effects. Among
these four, COV has smallest RMSE. For inference, asymptotic $95\%$
confidence intervals provide coverage very close to the nominal coverage
for GPSM, which confirms our inference theory in Web Appendix. Asymptotic
$95\%$ confidence intervals for GPSS and weighting are also fairly
accurate, but COV leads to undercoverage, consistent with the results
in \citet{AI:2006} on the bias of matching estimators with multiple
covariates.

In the second design with six treatment levels, we consider propensity
score design as $p(w\mid X_{i})=\exp(X_{i}^{T}\beta_{w})/\sum_{w'=1}^{6}\exp(X_{i}^{T}\beta_{w'})$,
where $\beta_{1}^{T}=(0,0,0,0,0,0,0)$, $\beta_{2}^{T}=0.4\times(0,1,1,2,1,1,1)$,
$\beta_{3}^{T}=0.6\times(0,1,1,1,1,1,-5)$, $\beta_{4}^{T}=0.8\times(0,1,1,1,1,1,5)$,
$\beta_{5}^{T}=1.0\times(0,1,1,1,-2,1,1)$, and $\beta_{6}^{T}=1.2\times(0,1,1,1,-2,-1,1)$.
The outcome design is $Y_{i}(w)=X_{i}^{T}\gamma_{w}$, with $\gamma_{1}^{T}=(-1.5,1,1,1,1,1,1)$,
$\gamma_{2}^{T}=(-3,2,3,1,2,2,2)$, $\gamma_{3}^{T}=(3,3,1,2,-1,-1,-4)$,
$\gamma_{4}^{T}=(2.5,4,1,2,-1,-1,-3)$, $\gamma_{5}^{T}=(2,5,1,2,-1,-1,-2)$,
and $\gamma_{6}^{T}=(1.5,6,1,2,-1,-1,-1)$ with $\eta_{i}\sim N(0,1)$.
The sample sizes are $N_{w}=1000$, for all $w$.

In Figure 1 we present the results for the estimators for the fifteen
average treatment effects. The simulation setup creates six treatment
groups with strong separation in covariate distributions, which makes
it fundamentally difficulty removing all biases in estimating fifteen
treatment effects simultaneously. Overall GPSM outperforms the other
methods in terms of bias and coverage rates, with the coverage rate
for nominal $95\%$ confidence intervals never going below $0.75$.
To assess the performance of the weighting estimator it is useful
to look at the weights that underly the estimator. Normalizing the
weights so that they average to $N_w$ for each of the treatment levels,
the maximum weight for units in each of the treatment levels is $16.9$
(treatment level one), $21.2$ (treatment level two), $50.0$ (treatment
level three), $64.9$ (treatment level four), $21.2$ (treatment level
five), and $185.1$ (treatment level six). Even in the three treatment
level case these maximum weights are substantial. There the maximum
weight for units in each of the treatment levels is $16.5$ (treatment
level one), $95.8$ (treatment level two), and $17.9$ (treatment
level three).

\begin{figure}
\centering{
\includegraphics[width=160mm]{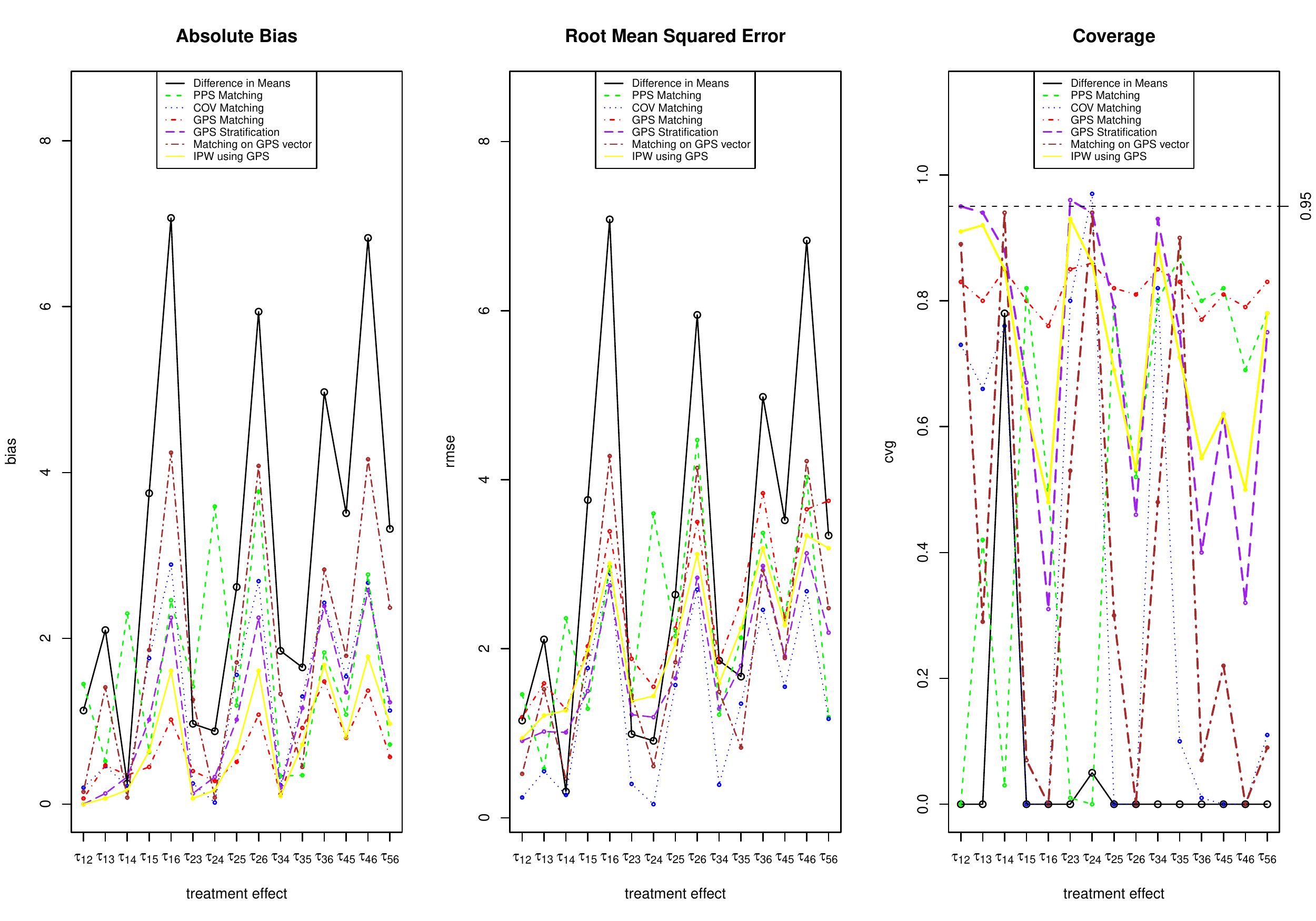}
}
\caption{Simulation Results, Design II.}
\label{fig:res2}
\end{figure}

In an extended simulation (see Web Appendix), we compare the performance
of the estimators under the combinations of (w/o) trimming and (correct/incorrect)
generalized propensity score model. When the propensity score model
is incorrect, the performance for all methods based on the propensity
score deteriorates. In particular, the weighting estimator shows huge
bias and variance and poor coverage for all parameters. GPSM is inferior
to COV in terms of bias and variance; however, it presents better
coverage for ten parameters out of fifteen parameters. This suggests
that when covariates are high dimensional, the inference for COV is
not satisfactory. After trimming, bias and variance are greatly reduced
and coverage is improved for all parameters for GPSM, GPSS and weighting,
which suggests that trimming can improve the performance of GPS based
methods.


\section{An Application}

We re-examined data from the REFLECTIONS (Real-world Examination of
Fibromyalgia: Longitudinal Evaluation of Costs and Treatments) study.
REFLECTIONS was a $12$-month prospective observational study of patients
being treated for fibromyalgia at $58$ outpatient sites in the US
and Puerto Rico. Patients had to be at least $18$ years of age and
initiating a new pharmacologic treatment for fibromyalgia. In keeping
with the observation nature of the study, inclusion and exclusion
criteria were kept to a minimum, no requirements on the nature of
the fibromyalgia treatment were made, and physicians decisions regarding
the proper treatment and care of patients were made in the course
of normal clinical practice. Data from patients were collected at
baseline during a standard office visit and at 1, $3,6$, and $12$
months post baseline via a computer assisted telephone interviews.
For details on the design of REFLECTIONS see \citet{R:2012}.

For this example, we focused on the analysis of three fibromyalgia
medication cohorts \citep{P:2015}: patients treated with an opioid
(either monotherapy or with other medications), patients treated with
tramadol but not an opioid, and patients not treated with tramadol
or an opioid (referred to as the Other cohort). The outcome variable
utilized here is the total score of Fibromyalgia Impact Questionnaire
(FIQ), which is composed of items measuring physical functioning,
number of days the patient felt well, number of days the patient felt
unable to work due to FM symptoms, and patient ratings of work difficulty,
pain intensity, fatigue, morning tiredness, stiffness, anxiety, and
depression. The total score ranges from $0$ to $80$ with lower scores
indicating better outcomes, and research suggests a $14\%$ reduction
(or $7.6$ points in this sample) is clinically relevant \citep{B:2009}.
The objective is to produce causal inference pairwise comparisons
between the cohorts all based on the same population (the population
represented by the trimmed sample), in order to test the study hypothesis
that there is no difference in FIQ total score among the three cohorts.

The generalized propensity scores (3 estimated probabilities for each
patient) were computed using a multinomial model with $32$ {predictors}
from demographics, baseline clinical characteristics, comorbidities,
resource use, prior fibromyalgia treatment, and physician information.

To help address lack of overlap of the populations, the modified Crump
\citep{C:2009} algorithm of Section 7 was applied. With the REFLECTIONS
data, $\lambda=29.88$, and thus patients were trimmed if their $\sum_{w=1}^{T}1/p(w|X_{i})$
value was greater than $43.52$. This resulted in removal of $363$
patients ($25\%$ of the sample), with $31(9\%)$ from the Opioid
cohort (OPI), $17(8\%)$ of the Tramadol cohort (TRA), and $315(34\%)$
from the Other cohort (OTH). Thus, the analysis cohort included $1101$
patients ($308$ OPI, $188$ TRA, $605$ OTH).

The trimming primarily removed patients in OTH who had high propensities
for being in OTH (and low propensities for either OPI or TRA) and
were under-represented in OPT and TRA. Using the trimmed sample, generalized
propensity score matching was implemented following the steps in Section
5 to produce counterfactual outcomes (imputed FIQ total scores) for
each patient and cohort. The quality of the matches appeared acceptable,
with the mean difference in propensity scores for the matched patients
ranging from $0.0012$ to $0.0014$ across the cohorts and the largest
matched pair with a difference of $0.035$.

The unadjusted mean changes(sd) from baseline to endpoint ($12$ months
post baseline) for the FIQ total score in the trimmed cohort were
$-2.4(12.3)$ for OPI, $-3.7(14.0)$ for TRA, and $-4.2(13.4)$ for
OTH, indicating small numerical improvement in pain symptoms. Table
\ref{table:tab4} summarizes the comparative analysis of the FIQ total
score improvement among the $3$ cohorts on the trimmed sample using
generalized propensity score matching (GPSM) and stratification (GPSS).
As a comparison, results using no bias adjustment (Difference in Means),
propensity score set matching (PSSM), weighting (W), and covariate
matching (COV) are included. Without bias adjustment, no cohort differences
reach the level of statistical significance, though OTH demonstrated
marginally greater reductions than OPI ($p=0.058$). Similarly, none
of the adjusted differences led to any statistical significant findings,
indicating similar health condition improvements at 12 months over
baseline among the three cohorts. Note that by using the same population
across all comparisons, pairwise differences across the cohorts using
the generalized propensity scoring methods (GPSM, GPSS) are consistent,
whereas using PPSM this is not true (the PPSM estimates suggest that
changing the treatment from OPI to TRA leads to an average effect
of -3.9, changing the treatment effect from TRA to OTH leads to an
average effect of -1.8, and changing the treatment from OPI to OTH
leads to an average effect of -1.4, which cannot all be true at the
same time). This illustrates the impact of differing populations can
have when using PPSM.

\begin{table}{\small
\centering
\caption{Analysis results: Change from Baseline to Endpoint (12 Months) for the FIQ total score. OPI = Opioid Cohort; TRA = Tramadol Cohort; OTH = Other Cohort.  TRA-OPI indicates the estimated difference in change from baseline scores for the Tramadol Cohort minus the same value for the Opioid cohort. Thus, negative values indicate greater reduction in symptoms for the first cohort. Analyses are on the Trimmed sample. Confidence Intervals were calculated using the same methods as for simulated study above.}
\begin{tabular}{l p{3cm}  p{3cm}  p{3cm} }
\hline
\multirow{2}{*}{ } & \multicolumn{3}{c}{Pairwise Differences Means ($95\%$ CI)} \\
Method		 & TRA - OPI	& OTH- OPI	 & OTH - TRA \\
\hline
{ DIF} & { -1.1 (-3.8, 1.1) } & { -1.7(-3.5, 0.3) } & { -0.6 (-2.6, 2.2) }\\
{ PPSM} & { -3.9 (-7.2, -0.6) }  & { -1.4 (-3.4, 0.6) }  & { -1.8 (-4.2, 0.6)} \\
{ PSSM} & { -2.5 (-6.6, 1.6) } & { -1.9 (-4.1, 0.4) } & { 0.7 (-3.1, 4.4) }\\
{ W} & { -0.8 (-5.4, 4.7)} & { -0.3 (-3.8, 5.1)} & { 0.4 (-2.3, 4.0)}\\
{ COV } & { -1.6 (-4.8, 1.5) } & { -1.5 (-3.8, 0.9) } & { 0.2 (-2.5, 2.8) }\\
{ GPSM} & { -1.6 (-4.3,1.1) } & { -0.9 (-2.8,1.1) } & { 0.7 (-1.8,3.2) }\\
{ GPSS} & { -1.6 (-5.5, 1.2) } & { -1.2 (-4.1, 0.9) } & { 0.4 (-2.2, 3.8)}\\
\hline
\end{tabular}
\label{table:tab4}
}
\end{table}


\section{Conclusion}
\label{sec:con}

In this paper, we develop new methods for estimating causal treatment effects using observational data in settings with multiple (more than two) treatment levels. 
Existing methods require additional assumptions assuming the existence of a scalar balance score, so as to facilitate matching and subclassification.
We show that, contrary to claims in the literature, matching and subclassification methods using the propensity score generalize naturally to the multi-level treatment case. We focus on matching and subclassification on the generalized propensity score using the notion of weak unconfoundedness, and show that adjusting for a scalar function of the covariates can remove all biases associated with differences in observed covariates.

As with other propensity based analyses, this approach 
{depends on correct specification of propensity score modeling, and}
does not resolve the potential for bias due to unmeasured confounding.  An initial simulation study and example demonstrated the potential benefits of the proposed approach at reducing bias and providing causal inference comparisons for multiple cohorts on a common population. 


\section*{Acknowledgment}
\label{sec:ack}

This study and the preparation of this paper were funded in full by Eli Lilly and Company, Indianapolis, IN, USA. Editing support was provided by Casie Polanco, inVentivHealth Clinical, Indianapolis, IN, USA. 

Declaration of personal interests:  Cui, Faries, and Kadziola are all employees and minor stockholders of Eli Lilly and Company. 

\section*{Supplementary Materials}

Supplementary Web Appendices, referenced in Sections 4 and 7, are available with this paper at
the \textit{Biometrics} website on Wiley Online Library. 
R code is available at \url{https://github.com/shuyang1987/multilevelMatching}.


\begin{thebibliography}{}

\bibitem[\protect\citeauthoryear{Abadie \& Imbens}{2006}]{AI:2006}
\textsc{ABADIE, A.} \& \textsc{IMBENS, G.} (2006).
 { Large sample properties of matching estimators for average treatment effects}.
 {\it  Econometrica} {\bf 74}, 235--67.

\bibitem[\protect\citeauthoryear{Abadie \& Imbens}{2012}]{AI:2012}
\textsc{ABADIE, A.} \& \textsc{IMBENS, G.} (2012).
 {A martingale representation for matching estimators}.
 {\it  J. Am. Statist. Assoc.} {\bf 107}, 833--43.


\bibitem[\protect\citeauthoryear{Bennett et al.}{2009}]{B:2009}
\textsc{BENNETT, R. M. },
\textsc{BUSHMAKIN, A. G.},
\textsc{CAPPELLERI, J. C.},
\textsc{ZLATEVA, G.}\&
\textsc{SADOSKY, A. B.}
(2009).
{Minimal clinically important difference in the fibromyalgia impact questionnairea}.
{\it The Journal of Rheumatology,} {\bf 36.6}, 1304--11.


\bibitem[\protect\citeauthoryear{Cadarette et al.}{2010}]{CA:2010}
\textsc{CADARETTE, S.},\textsc{GAGNE, J.}, \textsc{SOLOMON, D.},\textsc{KATZ, J.} \& \textsc{STURMER, T. } (2010).
 {Confounder summary scores when comparing the effects of multiple drug exposures}.
 {\it  Pharmacoepidemiol. Drug Saf.} {\bf 19}, 2--9.

\bibitem[\protect\citeauthoryear{Cattaneo}{2010}]{C:2010}
\textsc{CATTANEO, M.} (2010).
 {Efficient semiparametric estimation of multi-level treatment effects under ignorability}.
 {\it J. Econom.} {\bf 155}, 138--54.

\bibitem[\protect\citeauthoryear{Cochran}{1968}]{C:1968}
\textsc{COCHRAN, W.} (1968).
 {The effectiveness of adjustment by subclassification in removing bias in observational studies}.
 {\it Biometrics} {\bf 24}, 295--314.

\bibitem[\protect\citeauthoryear{Cole \& Frangakis}{2009}]{CF:2009}
\textsc{COLE, S.} \& \textsc{FRANGAKIS, C.} (2009).
 {The consistency assumption in causal inference: a definition or an assumption?}.
 {\it Epidemiology} {\bf 20}, 3--5.

\bibitem[\protect\citeauthoryear{Crump et al.}{2009}]{C:2009}
\textsc{CRUMP, R.}, \textsc{HOTZ, V. J.}, \textsc{IMBENS, G.} \& \textsc{MITNIK, O.} (2009).
 {Dealing with limited overlap in estimation of average treatment effects}.
 {\it Biometrika} {\bf 96}, 187--99.

\bibitem[\protect\citeauthoryear{Dawid}{1979}]{D:1979}
\textsc{DAWID, A.P.} (1979).
 {Some misleading arguments involving conditional independence}.
 {\it J. Roy. Statist. Soc. B.} {\bf 41}, 249--52.


\bibitem[\protect\citeauthoryear{Frolich}{2004a}]{F:2004}
\textsc{FR{\"O}LICH, M.} (2004a).
 {Finite-Sample Properties of Propensity-Score Matching and Weighting Estimators}.
 {\it Rev. Econ. Stat.} {\bf 86}, 77-90.

\bibitem[\protect\citeauthoryear{Fr{\"o}lich}{2004b}]{F:2004b}
\textsc{FR{\"O}LICH, M.} (2004b).
{Programme evaluation with multiple treatments}. 
{\it Journal of Economic Surveys,} {\bf 18.2} 181--224.

\bibitem[\protect\citeauthoryear{Foster}{2003}]{F:2003}
\textsc{FOSTER, E. M.} (2003).
{Propensity score matching: an illustrative analysis of dose response}.
{\it Medical care,} {\bf 41.10} 1183--1192.

\bibitem[\protect\citeauthoryear{Lee, Lessler, \& Stuart}{2010}]{L:2010}
\textsc{LEE, B. K.}, \textsc{LESSKER, J.} \& \textsc{STUART, E.A.} (2010).
{Improving propensity score weighting using machine learning}.
{\it Statistics in medicine} {\bf 29.3} 337--346.

\bibitem[\protect\citeauthoryear{Guo \& Fraser}{2010}]{GF:2010}
\textsc{GUO, S.} \& \textsc{FRASER M.} (2010).
 {\it Propensity Score Analysis}, Sage.

\bibitem[\protect\citeauthoryear{Hahn}{1998}]{H:1998}
\textsc{HAHN, J.} (1998).
 {On the role of the propensity score in efficient semiparametric estimation of average treatment effects}.
 {\it Econometrica} {\bf 66}, 315--31.





\bibitem[\protect\citeauthoryear{Hirano \& Imbens}{2004}]{HI:2004}
\textsc{HIRANO, K.} \& \textsc{IMBENS, G.}  (2004).
 {The Propensity Score with Continuous Treatments}.
{\it Applied Bayesian Modelling and Causal Inference from Missing
Data Perspectives}, Gelman and Meng (eds.), Wiley.


\bibitem[\protect\citeauthoryear{Hirano, Imbens \& Ridder}{2003}]{HIR:2003}
\textsc{HIRANO, K.}, \textsc{IMBENS, G.} \& \textsc{RIDDER, G.} (2003).
 {Efficient estimation of average treatment effects using the estimated propensity score}.
 {\it Econometrica} {\bf 74}, 1161--89.

\bibitem[\protect\citeauthoryear{Huber, Lechner \& Wunsch}{2013}]{H:2013}
\textsc{HUBER, M.}, \textsc{LECHNER, M.} \& \textsc{WUNSCH, C.} (2013).
{The performance of estimators based on the propensity score}.
{\it Journal of Econometrics,} {\bf 175.1} 1--21.

\bibitem[\protect\citeauthoryear{Imai \& Van Dyk}{2004}]{IV:2004}
\textsc{IMAI, K.} \& \textsc{VAN DYK, D.} (2004).
 {Causal inference with general treatment regimes: generalizing the propensity score}.
 {\it J. Am. Statist. Assoc.} {\bf 99}, 854--66.


\bibitem[\protect\citeauthoryear{Imai \& Ratkovic}{2014}]{Im:2014}
\textsc{IMAI, K.} \& \textsc{RATKOVIC, M.} (2014). 
{Covariate Balancing Propensity Score}. 
{\it Journal of the Royal Statistical Society, Series B (Statistical Methodology)} {\bf 76(1)},  243--246.

\bibitem[\protect\citeauthoryear{Imbens}{2000}]{I:2000}
\textsc{IMBENS, G.} (2000).
 {The role of the propensity score in estimating dose-response functions}.
 {\it Biometrika} {\bf 87}, 706--10.

\bibitem[\protect\citeauthoryear{Imbens}{2004}]{I:2004}
\textsc{IMBENS, G.} (2004).
 {Nonparametric Estimation of Average Treatment Effects Under Exogeneity: A Review}.
 {\it Rev. Econ. Stat.} {\bf 86}, 4--29.

\bibitem[\protect\citeauthoryear{Imbens \& Rubin}{2015}]{IR:2015}
\textsc{IMBENS, G.} \& \textsc{RUBIN, D.}(2015).
 {\it An Introduction to Causal inference in the Statistical, Biomedical and Social Sciences}.
 {Cambridge: Cambridge University Press}.

\bibitem[\protect\citeauthoryear{Joffe \& Rosenbaum}{1999}]{Jo:1999}
\textsc{JOFFE, M. M.} \& \textsc{ROSENBAUM, P. R.} (1999)
{Invited commentary: propensity scores}.
{\it American Journal of Epidemiology,} {\bf 150.4}  327--333.


\bibitem[\protect\citeauthoryear{Kang \& Schafer}{2007}]{KS:2007}
\textsc{KANG, J.} \& \textsc{SCHAFER, J.}(2007).
{Demystifying Double Robustness:
A Comparison of Alternative Strategies for
Estimating a Population Mean from
Incomplete Data}.
 {\it Statistical Science}.
{\bf 22}(4), 523--539.


\bibitem[\protect\citeauthoryear{Lechner}{2001}]{L:2001}
\textsc{LECHNER, M.} (2001).
 {Identification and Estimation of Causal Effects of Multiple Treatments under the Conditional Independence Assumption}.
 {\it In Econometric Evaluations of Active Labor Market Policies in Europe, M. Lechner \& F. Pfeiereds. Heidelberg:Physica} pp. 43--58.





\bibitem[\protect\citeauthoryear{Lu et al.}{2001}]{Lu:2001}
\textsc{LU, B.}, \textsc{ZANUTTO, E.}, \textsc{HORNIK, R.} \& \textsc{ROSENBAUM, P. R} (2001).
{Matching with doses in an observational study of a media campaign against drug abuse}. 
{\it Journal of the American Statistical Association} {\bf 96.456} 1245--1253.

\bibitem[\protect\citeauthoryear{McCaffrey et al.}{2013}]{M:2013}
\textsc{MCCAFFREY, D.F.}, \textsc{GRIFFIN,B.A.}, \textsc{ALMIRALL,D.}, \textsc{SLAUGHTER, M.E.}, \textsc{RAMCHAND, R.} \& \textsc{BURGETTEB, L.F.}(2013).
 {A tutorial on propensity score estimation for multiple treatments using generalized boosted models}.
 {\it Stat. Med.} {\bf 32}, 3388--414.

\bibitem[\protect\citeauthoryear{Morgan \& Winship}{2007}]{MW:2007}
\textsc{MORGAN, S.} \& \textsc{WINSHIP,C.}(2007).
 {\it Counterfactuals and Causal Inference}.
 {Cambridge: Cambridge University Press}.

\bibitem[\protect\citeauthoryear{Peng et al.}{2015}]{P:2015}
\textsc{PENG, X.},
\textsc{ROBINSON, R. L.},
\textsc{MEASE, P. },
\textsc{KROENKE, K. },
\textsc{WILLIAMS, D. A.},
\textsc{CHEN, Y.},
\textsc{FARIES, D.},
\textsc{WOHLREICH, M.},
\textsc{MCCARBERG, B.} \&
\textsc{HANN, D.}(2015).
{Long-term evaluation of opioid treatment in fibromyalgia}.
{\it The Clinical journal of pain,} {\bf 31.1}, 7--13.


\bibitem[\protect\citeauthoryear{Rassen et al.}{2013}]{R:2013}
\textsc{RASSEN, J.}, \textsc{SHELAT, A.}, \textsc{FRANKLIN, J.}, \textsc{GLYNN, R.} \& \textsc{SCHNEEWEISS, S.}(2013).
 {Matching by propensity score in cohort studies with three treatment groups}.
 {\it Epidemiology} {\bf 24}, 401--9.





\bibitem[\protect\citeauthoryear{Rubin}{1974}]{R:1974}
\textsc{RUBIN, D.}(1974).
 {Estimating causal effects of treatments in randomized and nonrandomized study}.
 {\it J. Educ. Psychol.} {\bf 65}, 688--701.


\bibitem[\protect\citeauthoryear{Rubin}{1978}]{R:1978}
\textsc{RUBIN, D.}(1978).
 {Bayesian inference for causal effects: The Role of Randomization}.
 {\it Ann. Stat.} {\bf 6}, 34--58.



\bibitem[\protect\citeauthoryear{Robins, et al.}{2000}]{R:2000}
\textsc{ROBINS, J. M.}, \textsc{HERNAN, M. A.} \& \textsc{BRUMBACK, B.} (2000). 
{Marginal structural models and causal inference in epidemiology}. 
{\it Epidemiology,} {\bf 11(5)}, 550-560.

\bibitem[\protect\citeauthoryear{Robinson et al.}{2012}]{R:2012}
\textsc{ROBINSON, R.L.}, \textsc{KROENKE, K.}, \textsc{MEASE, P.}, \textsc{WILLIAMS, D. A.}, \textsc{CHEN, Y.}, \textsc{D'SOUZA, D.}, \textsc{WOHLREICH, M.} \& \textsc{MCCARBERG, B.}(2012).
 {Burden of Illness and Treatment Patterns for Patients with Fibromyalgia}.
 {\it Pain Med.} {\bf 13}, 1366--76.

\bibitem[\protect\citeauthoryear{Rosenbaum \& Rubin}{1983}]{RR:1983}
\textsc{ROSENBAUM, P.} \& \textsc{RUBIN, D.} (1983).
 {The Central Role of the Propensity Score in Observational Studies for Causal Effects}.
 {\it Biometrika} {\bf 70}, 41--55.

\bibitem[\protect\citeauthoryear{Setoguchi et al.}{2008}]{R:2008}
\textsc{SETOGUCHI, SOKO, ET AL.} (2008).
{Evaluating uses of data mining techniques in propensity score estimation: a simulation study}.
{\it Pharmacoepidemiology and drug safety,} {\bf 17.6} 546.


\bibitem[\protect\citeauthoryear{Zanutto, Lu, \& Hornik}{2005}]{Z:2005}
\textsc{ZANUTTO, E.}, \textsc{LU, B.} \& \textsc{HIRNIK, R.} (2005).
{Using propensity score subclassification for multiple treatment doses to evaluate a national antidrug media campaign}.
{\it Journal of Educational and Behavioral Statistics,} {\bf 30.1}, 59--73.



\end{thebibliography}

%
%
%

\end{document}